# Research the influence of road surface conditions on vibration and safety of Semi Trailer vehicle dynamics

Phat Khau Tan[a,✉], Danh Nguyen Thanh[a]

[a]*Vinh Long University of Technology Education, Vietnam*

Abstract

**Purpose:** Depending on the road surface profile, moving speed, transport weight, etc., the vehicle's body acceleration and dynamic load coefficient change as it moves. This study's goal is to ascertain the result of the road surface's influence on the vehicle, calculate the average vehicle body displacement acceleration and dynamic load coefficient, and validate your findings through testing. To confirm the simulation model's accuracy, experimental and theoretical simulation results are compared..

**Methods:** The Neuton-Euler approach and the multi-body system structure separation method are used in the paper to create dynamic equations using theoretical research methods [1]. Simulations are carried out using Matlab Simulink software to examine and assess the dynamics.

**Results:** According to ISO 8608:2016 standard [2], theoretical research results have determined the average acceleration of the vehicle body, dynamic load coefficient, horizontal sway angle, and longitudinal sway angle when the vehicle moves on profiles C, D , E, F. Experiments show that there is an error of 7.4% and 8.1% when measuring the average acceleration $\ddot{Z}_{a\max}$ and dynamic load coefficient $k_{đmax}$ when the vehicle moves at a speed of 50 km/h on the road surface C compared to theory.

1. Introduction

Today, many semi-trailers are circulating in Vietnam that follow foreign road standards. Due to poor road conditions, a year-round rainy climate, and frequent overloading, semi-trucks in Vietnam frequently lose stability and result in traffic accidents. Huge, hefty trucks that are used to transport bulky, huge items are called semi-trailers. When engaging in traffic, there are numerous possible risk factors for extensive activities. However, the fleet of semi-trailers has a very heavy load on the road, which has a direct impact on the road surface's longevity.

Starting from the above problems, the authors built a dynamic model and wrote a system of equations for the dynamics of semi-trailers[3]; Set up a simulation model to investigate the influence of road surface conditions on dynamic load and semi-trailer body acceleration as an input basis for calculation problems to determine vehicle durability and conduct experimental testing prove the accuracy of the theoretical model.

Semi-trailer vehicles work with heavy loads on relatively poor road profiles in Vietnam, so they often cause large dynamic loads that affect structural durability, cargo safety, and dynamic safety of cars and destroy the roadbed. Evaluation statistics on vehicle body acceleration include:

Oscillation acceleration is an important parameter to evaluate the smoothness of motion, it takes into account the simultaneous influence of amplitude and frequency of oscillation. Because damped free oscillation only lasts for a few cycles, determining the oscillation acceleration will be of great significance when studying forced oscillation with road surface excitation.

- For humans [4],[5]: The limit acceleration value in the OX (vehicle longitudinal direction), OY (vehicle horizontal direction), OZ (vertical direction) directions are determined experimentally as follows: $\ddot{Z}_x < 1,0$ $m/s^2$; $\ddot{Z}_y < 0,7$ $m/s^2$; $\ddot{Z}_z < 2,5$ $m/s^2$ !

- For goods [6],[7]: According to the Standards of the German Packaging Association, extreme vertical acceleration values are used to evaluate the smoothness of goods:

$\ddot{Z}_{a\max} = 3 m/s^2$: Warning limit, threshold for planning vehicle or road repairs;

$\ddot{Z}_{a\max} = 5 m/s^2$: Intervention limit, threshold for immediate road or vehicle repair.

In addition, the author uses dynamic load assessment criteria to evaluate the dynamic safety of cars such as:

The load coefficient ($k_d$) is used to evaluate the influence of the road on the wheel reaction $F_z$, defined as follows [8], [9], [10]:

$$k_d = \frac{F_{zz}}{F_{zt}}; \quad F_{zz} = F_{zt} + F_{zd} \qquad (1)$$

There are several variables that are present: load factor ($k_d$); vertical wheel reaction force ($F_{zz}$); vertical wheel response force in the static state ($F_{zt}$); and vertical dynamic load ($F_{zd}$) (N).

To evaluate a car from the point of view of motion safety, it is necessary to determine the ratio between dynamic and static wheel loads $F_{zd}(t)/F_{zt}$.

The dynamic load $F_{zd}$ always changes over time, so we use the mean square value of the dynamic load RMS ($F_{zd}$).

In addition, when the car vibrates, people are concerned about the tire's grip on the road surface. The oscillating car satisfies the criteria for smoothness, but the wheels have poor grip on the road, causing a loss of stability when driving the vehicle. Therefore, it is possible to use the mean square value of the relative displacement between the wheels and the road surface roughness to evaluate the adhesion (contact) of the wheels with the road surface.

Therefore, the motion safety index for cars can be evaluated according to the following mathematical expression [8], [9], [10]:

$$RMS(F_zd) = \frac{1}{4}\left(\sum_{i=1}^{4}\sqrt{\frac{1}{T}\int_0^T F_{zdi}^2 dt}\right) \ [N] \qquad (2)$$

When it is necessary to evaluate the dynamic load interaction between the vehicle and the road, $k_{đmin}$ and $k_{đmax}$ can be used, which are defined as follows [8], [9]:

$$k_{đ\min} = \frac{\min(F_{zd})}{F_{zt}}; \quad 0 \leq k_{đ\min} \leq 1 \tag{3}$$

Some limitations for $k_{đmin}$ are as follows [8], [9]:

- $k_{đmin}$ = 0.5: warning limit;

- $k_{đmin}$ = 0: intervention limit, cake separation occurs, ($F_z$ = 0).

$$k_{đ\max} = \frac{\max(F_{zd})}{F_{zt}} \tag{4}$$

For random agitation, $k_{đmax}$ in formula (1.5) is calculated as follows:

$$k_{đ\max} = \frac{RMS(F_{zđ})}{F_{zt}} \tag{5}$$

According to some research works [9], $k_{đmax} \leq 1.5$ is the limit that ensures the durability of vehicle details.

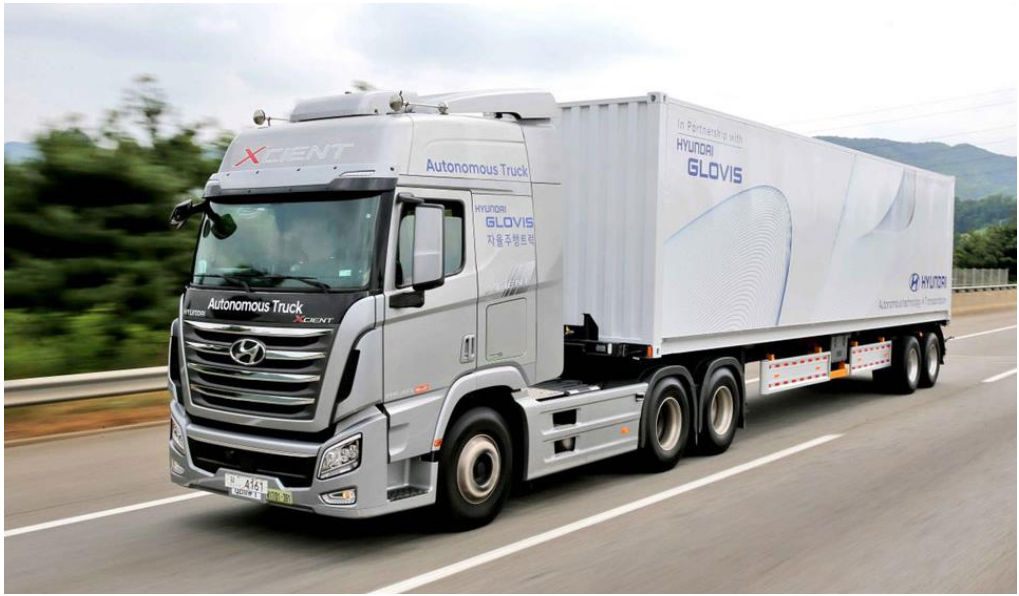

Figure 1. Convoy of semi-trailers

**Dynamic model of semi-trailer vehicle**

The semi-trailer fleet has a large mass and size, a complex multi-body structure with geometric and physical nonlinear connections. The structure is divided into two large masses: the front tractor mass and the rear trailer mass with a balanced suspension system on the axles with high nonlinear physical coefficients (Figure 1).

This study's goal is to ascertain the result of the road surface's influence on the vehicle, calculate the average vehicle body displacement acceleration and dynamic load coefficient, and validate your findings through testing.

The dynamic model of a semi-trailer fleet is very complex, To simplify the process of setting up the model we use some assumptions as follows:

(1). Longitudinal axially symmetric semi-trailer fleet;

(2). The vehicle axle does not rotate around the y-axis ($\varphi_{Ai}$ = 0, i=1÷5); The vehicle axle moves forward along the x-axis and rotates around the z-axis along with the suspended mass of the tractor or trailer semi-trailer ($x_{Ai}$ = $x_{c1}$, $\psi_{Ai} = \psi_{c1}$, i=1÷3); ($x_{Ai}$ = $x_{c2}$, $\psi_{Ai} = \psi_{c2}$, i=5÷6);

(3). Neglect elasticity and friction in the saddle joint, horizontal air resistance, and wheel torque around the z-axis;

(4). The elasticity of the suspension system is linear in the working domain and nonlinear when reaching the travel limit lug;

(5). The radial stiffness of the tire does not change without separating the tire.

Dynamic model of semi-trailer fleet (Figure 3) when moving in the road plane depends on longitudinal and lateral forces $F_{xij}$ and $F_{yij}$. These forces depend on the reaction force $F_{zij}$. Therefore, the author writes the kinematic equations into the following two groups:

(8) Dynamic equation of the semi-trailer convoy in the road plane (OXY) to calculate the movement of the convoy in the x and y directions and the rotation of the vehicle body around the z-axis.

(ii) Dynamic equation of semi-trailer fleet in the vertical plane to determine the reaction force $F_z$.

$$(m_{c1} + \sum_{1}^{3} m_{Ai})(\ddot{x}_{c1} - \dot{\psi}_{c1}\dot{y}_{c1}) = (F_{x11} - F_{R11})\cos\delta_{11} + (F_{x12} - F_{R12})\cos\delta_{12} - F_{kx1} - F_{y11}\sin\delta_{11}$$
$$- F_{y12}\sin\delta_{12} - (F_{R21} + F_{R22} + F_{R31} + F_{R32}) + (F_{x21} + F_{x22} + F_{x31} + F_{x32}) \quad (6)$$

$$(m_{c1} + \sum_{1}^{3} m_{Ai})(\ddot{y}_{c1} + \dot{\psi}_{c1}\dot{x}_{c1}) = (F_{x11} - F_{R11})\sin\delta_{11} + (F_{x12} - F_{R12})\sin\delta_{12} + F_{y11}\cos\delta_{11} + F_{y12}\cos\delta_{12}$$
$$+ (F_{y21} + F_{y22} + F_{y31} + F_{y32}) - F_{ky1} \quad (7)$$

$$J_{zc1}\ddot{\psi}_{c1} = (F_{x11}\sin\delta_{11} + F_{x12}\sin\delta_{12} + F_{y11}\cos\delta_{11} + F_{y12}\cos\delta_{12})l_1 + (F_{x12}\cos\delta_{12} - F_{x11}\cos\delta_{11} +$$
$$F_{y11}\sin\delta_{11} - F_{y12}\sin\delta_{12})b_1 + (F_{x22} - F_{x21})b_2 + (F_{x32} - F_{x31})b_3 - (F_{y21} + F_{y22})l_2 - (F_{y31} + F_{y32})l_3 \quad (8)$$

$$(m_{c2}+\sum_{4}^{6}m_{Ai})(\ddot{x}_{c2}-\dot{\psi}_{c2}\dot{y}_{c2})=F_{x41}+F_{x42}+(F_{x51}+F_{x52})+F_{x61}+F_{x62}+F_{kx2} \quad (9)$$

$$(m_{c2}+\sum_{4}^{6}m_{Ai})(\ddot{y}_{c2}+\dot{\psi}_{c2}\dot{x}_{c2})=F_{ky2}+F_{y41}+F_{y42}+F_{y51}+F_{y52} \quad (10)$$

$$J_{zc2}\ddot{\psi}_{c2}=(F_{x42}-F_{x41})b_4+(F_{x52}-F_{x51})b_5-(F_{y41}+F_{y42})l_4-(F_{y51}+F_{y52})l_5 \quad (11)$$

$$m_{c1}(\ddot{z}_{c1}-\dot{\varphi}_{c1}\dot{x}_{c1})=F_{C11}+F_{C12}+F_{C21}+F_{C22}+F_{C31}+F_{C32} \quad (12)$$

$$J_{yc1}\ddot{\varphi}_{c1}=(F_{C11}+F_{C12})l_1-(F_{C21}+F_{C22})l_2-(F_{C31}+F_{C32})l_3+(F'_{x11}+F'_{x12})(h_{c1}-r_1)$$
$$+(F'_{x21}+F'_{x22})(h_{c1}-r_2)+(F'_{x31}+F'_{x32})(h_{c1}-r_3) \quad (13)$$

$$m_{c2}(\ddot{z}_{c2}-\dot{\varphi}_{c2}\dot{x}_{c2})=F_{C41}+F_{C42}+F_{C51}+F_{C52} \quad (14)$$

$$J_{yc2}\ddot{\varphi}_{c2}=-(F_{C41}+F_{C42})l_4-(F_{C51}+F_{C52})l_5+(F'_{x41}+F'_{x42})(h_{c2}-r_4)+(F'_{x51}+F'_{x52})(h_{c2}-r_5)$$
$$+F_{kx2}(h_{c2}-h_{k2}) \quad (15)$$

$$J_{xc1}\ddot{\beta}_{c1}=\sum_{i=1}^{i=3}(F_{Ci2}+F_{Ki2}-F_{Ci1}-F_{Ki1})w_i+\sum_{i=1}^{i=3}F_i(h_{c1}-h_{Bi})+M_{kx1} \quad (16)$$

$$m_{A1}(\ddot{z}_{A1}+\dot{\beta}_{A1}\dot{y}_{A1})=F_{CL11}+F_{KL11}+F_{CL12}+F_{KL12}-F_{C11}-F_{K11}-F_{C12}-F_{K12} \quad (17)$$

$$m_{A1}(\ddot{y}_{A1}-\dot{\beta}_{A1}\dot{z}_{A1})=F_1+F_{y11}+F_{y12} \quad (18)$$

$$J_{Ax1}\ddot{\beta}_{A1}=(F_{C11}+F_{K11}-F_{C12}-F_{K12})w_1+(F_{CL12}+F_{KL12}-F_{CL11}-F_{KL11})b_1-F_{y11}(r_{11}+\xi_{A11})$$
$$-F_{y12}(r_{12}+\xi_{A12})+F_1(h_{B1}-r_1) \quad (19)$$

$$m_{A2}(\ddot{z}_{A2}+\dot{\beta}_{A2}\dot{y}_{A2})=F_{CL21}+F_{KL21}+F_{CL22}+F_{KL22}-F_{C21}-F_{K21}-F_{C22}-F_{K22} \quad (20)$$

$$m_{A2}(\ddot{y}_{A2}-\dot{\beta}_{A2}\dot{z}_{A2})=F_2+F_{y21}+F_{y22} \quad (21)$$

$$J_{Ax2}\ddot{\beta}_{A2}=(F_{C21}+F_{K21}-F_{C22}-F_{K22})w_2+(F_{CL22}+F_{KL22}-F_{CL21}-F_{KL21})b_2-F_{y21}(r_{21}+\xi_{A21})$$
$$-F_{y22}(r_{22}+\xi_{A22})+F_2(h_{B2}-r_2) \quad (22)$$

$$m_{A3}(\ddot{z}_{A3}+\dot{\beta}_{A3}\dot{y}_{A3})=F_{CL31}+F_{KL31}+F_{CL32}+F_{KL32}-F_{C31}-F_{K31}-F_{C32}-F_{K32} \quad (23)$$

$$m_{A3}(\ddot{y}_{A3}-\dot{\beta}_{A3}\dot{z}_{A3})=F_3+F_{y31}+F_{y32} \quad (24)$$

$$J_{Ax3}\ddot{\beta}_{A3}=(F_{C31}+F_{K31}-F_{C32}-F_{K32})w_3+(F_{CL32}+F_{KL32}-F_{CL31}-F_{KL31})b_3$$
$$-F_{y31}(r_{31}+\xi_{A31})-F_{y32}(r_{32}+\xi_{A32})+F_3(h_{B3}-r_3) \quad (25)$$

$$J_{xc2}\ddot{\beta}_{c2} = \sum_{i=1}^{i=3}(F_{Ci2}+F_{Ki2}-F_{Ci1}-F_{Ki1})w_i + \sum_{i=1}^{i=3} F_i(h_{c2}-h_{Bi}) - M_{kx2} \quad (26)$$

$$m_{A4}(\ddot{z}_{A4}+\dot{\beta}_{A4}\dot{y}_{A4}) = F_{CL41}+F_{KL41}+F_{CL42}+F_{KL42}-F_{C41}-F_{K41}-F_{C42}-F_{K42} \quad (27)$$

$$m_{A4}(\ddot{y}_{A4}-\dot{\beta}_{A4}\dot{z}_{A4}) = F_4 + F_{y41} + F_{y42} \quad (28)$$

$$J_{Ax4}\ddot{\beta}_{A4} = (F_{C41}+F_{K41}-F_{C42}-F_{K42})w_4 + (F_{CL42}+F_{KL42}-F_{CL41}-F_{KL41})b_4 - F_{y41}(r_{41}+\xi_{A41})$$
$$-F_{y42}(r_{42}+\xi_{A42}) + F_4(h_{B4}-r_4) \quad (29)$$

$$m_{A5}(\ddot{z}_{A5}+\dot{\beta}_{A5}\dot{y}_{A5}) = F_{CL51}+F_{KL51}+F_{CL52}+F_{KL52}-F_{C51}-F_{K51}-F_{C52}-F_{K52} \quad (30)$$

$$m_{A5}(\ddot{y}_{A5}-\dot{\beta}_{A5}\dot{z}_{A5}) = F_5 + F_{y51} + F_{y52} \quad (31)$$

$$J_{Ax5}\ddot{\beta}_{A5} = (F_{C51}+F_{K51}-F_{C52}-F_{K52})w_5 + (F_{CL52}+F_{KL52}-F_{CL51}-F_{KL51})b_5$$
$$-F_{y51}(r_{51}+\xi_{A51}) - F_{y52}(r_{52}+\xi_{A52}) + F_5(h_{B5}-r_5) \quad (32)$$

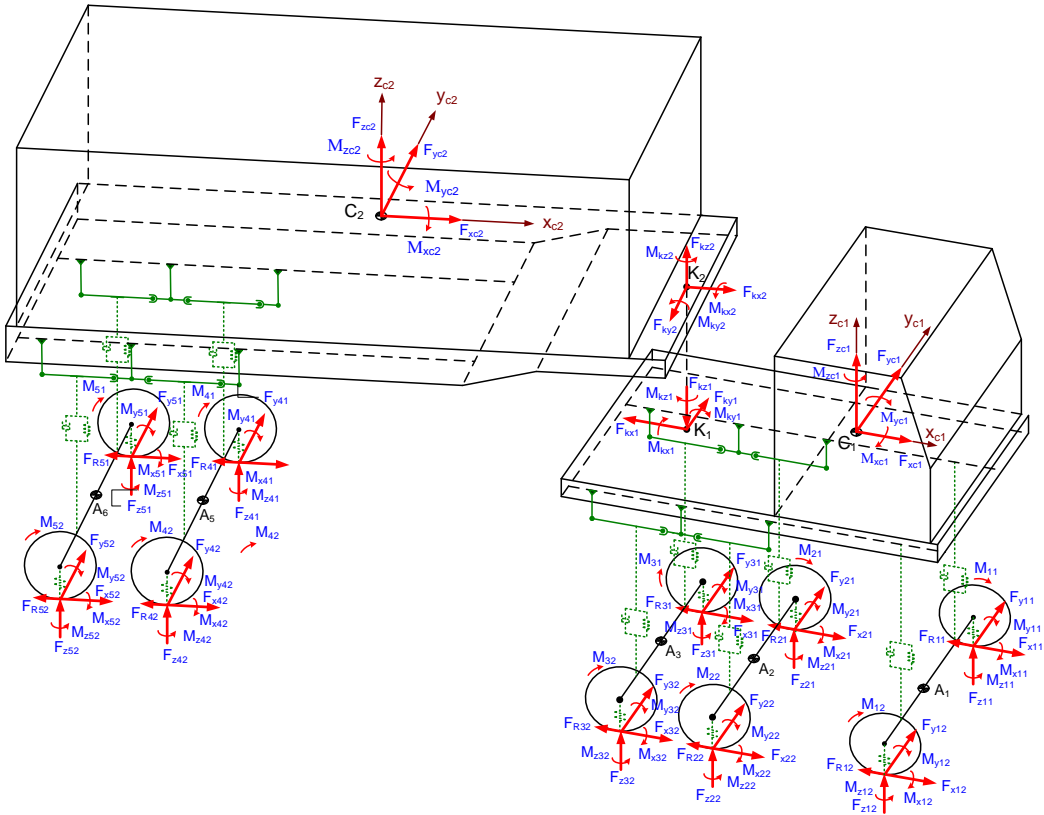

Figure 2. Diagram of forces acting on a semi-trailer convoy

Index: l₁,l₂,l₃ distance from bridge 1,2,3 to tractor truck center; l₄,l₅ distance from bridge 4,5 to semi-trailer center of gravity; $2b_i$, $2w_i$ he i wheel track and the distance between the i bridge spring pins; $\varphi_{c1}, \beta_{c1}, \psi_{c1}$ Angle of rotation of the tractor's body around the axis $y_{c1}, x_{c1}, z_{c1}$; $\varphi_{c2}, \beta_{c2}, \psi_{c2}$ rotation angle of the semi-trailer body around the axis $\varphi_{c2}, \beta_{c2}, \psi_{c2}$; $z_{c1}, \ddot{z}_{c1}$ displacement, vertical acceleration of tractor truck center of gravity; $x_{c1}, \dot{x}_{c1}, \ddot{x}_{c1}$ displacement, velocity, acceleration along of tractor truck center of gravity; $y_{c1}, \dot{y}_{c1}, \ddot{y}_{c1}$ displacement, velocity, horizontal acceleration of tractor truck center of gravity; $z_{c2}, \ddot{z}_{c2}$ displacement and vertical acceleration of semi-trailer center of gravity; $x_{c2}, \dot{x}_{c2}, \ddot{x}_{c2}$ displacement, velocity, acceleration along the center of gravity of the semi-trailer; $y_{c2}, \dot{y}_{c2}, \ddot{y}_{c2}$ displacement, velocity, horizontal acceleration of semi-trailer center of gravity; $J_{y11}$, $J_{y12}$ y-axis moment of inertia vertical stabilizer bar No. 1 of left and right suspension system; $F_{Cij}$ elastic force of wheel suspension system ij; $F_{Kij}$ the resistance of the ij wheel suspension; $F_{Clij}$ radial elastic force of tire ij.

2. Survey to determine vibrations and loads

**Survey to calculate dynamic load and acceleration using theoretical techniques**

When conducting simulations, survey conditions are carried out in the case of vehicles moving on road profiles A, B, C, and D according to ISO 8608:2016 [2] standards with a survey speed of 50 km/h, this is also the maximum allowable speed range when traveling within city limits in Vietnam.

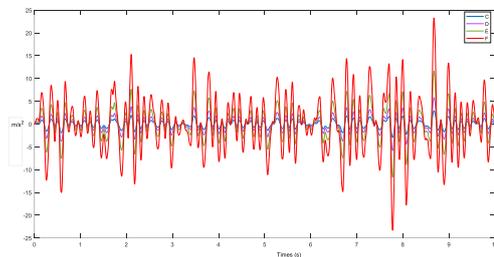

Figure 3. Vehicle body acceleration graph

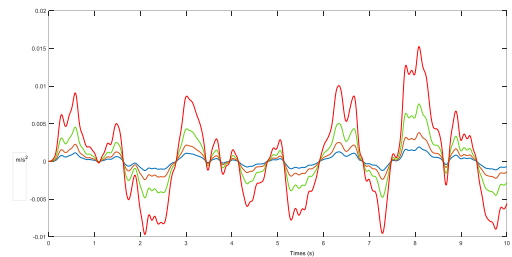

Figure 4. Vehicle body displacement diagram

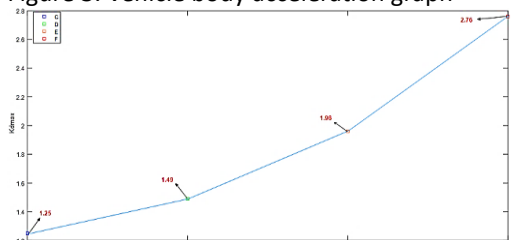

Figure 5. Dynamic load coefficient $k_{đmax}$

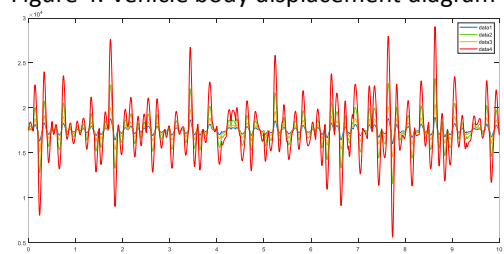

Figure 6. Dynamic load $F_{z1}$ acting on bridge 1

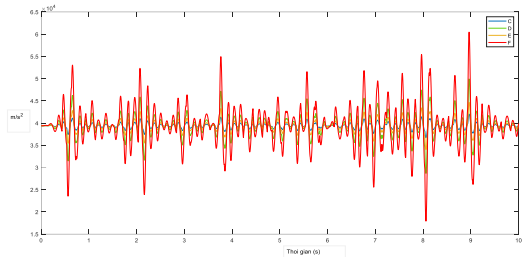
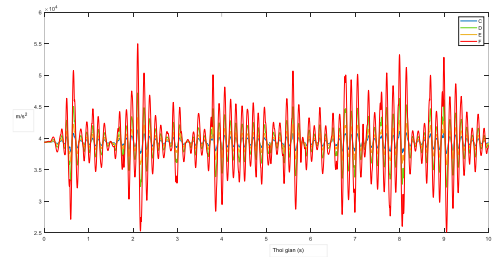

Figure 7. Dynamic load $F_{z2}$ acting on bridge 2     Figure 8. Dynamic load $F_{z3}$ acting on bridge 3

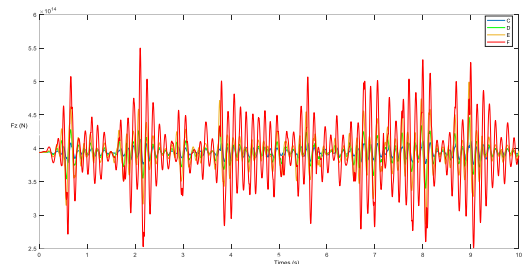
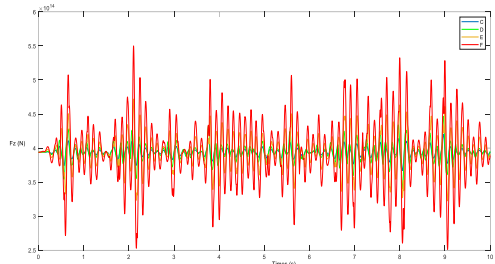

Figure 9. Dynamic load $F_{z4}$ acting on bridge 4     Figure 10. Dynamic load $F_{z5}$ acting on bridge 5

The survey results are shown in Figure 3 for pavements C, D, E, F. Vehicles move most smoothly (vertical acceleration does not increase suddenly) at pavements C and D, where the acceleration The average vehicle body is $\ddot{Z}_{a\max}$ = [2.3655, 2.8795] m/s$^2$ < 3 m/s$^2$ (Cargo safety warning limit); At road surfaces E and F, the average acceleration of the vehicle body is $\ddot{Z}_{a\max}$ = [7.2; 12.4] m/s$^2$ > 3 m/s$^2$ (exceeds Warning limits, vehicle or road repair planning threshold) [6],[7].

Through the survey results in Figure 5, when the vehicle moves at speeds of 50 km/h on road surface profiles C, D, E, and F according to ISO 8608:2016 standard, the value of dynamic load coefficient $k_{đmax}$ when Vehicles moving on road profiles C and D are $k_{đmax}$ = [1.25; 1.49] are all within the limit to ensure the durability of vehicle details $k_{đmax}$ 1.5 [9]. With road profiles E and F, the dynamic load value $k_{đmax}$ is respectively $k_{đmax}$ = [1.96; 2.76] all exceed the warning limit for the durability of car parts $k_{đmax}$ [1.96; 2.79] > 1.5 [9]. At a travel speed of 50 km/h, the two road profiles C and D still ensure dynamic safety and vehicle longevity; In addition, the two road profiles E and F, do not choose a speed of 50 km/h when traveling because it will exceed the dynamic safety warning limit and the durability of car parts.

Figures 6, 7, 8, 9, 10 show the results of the dynamic load survey on axles 1,2,3,4,5 of the semi-trailer fleet. The survey chart shows that the dynamic load value $F_z$ increases as the vehicle moves on worse road types with the same speed of 50 km/h on road profiles C, D, E, and F according to ISO 8608 standards: 2016 [2].

3. Experiment to verify the theoretical model

The measurement method to determine the normal reaction force from the road surface to the wheel is using the measurement method using a loadcell force sensor placed at the wheel where there is direct contact between the wheel and the axle. The vehicle body acceleration sensor 8152C from Kistler is located at the center of the vehicle's body and sends signals to the computer. Besides, speed sensors are also installed on the wheels to determine the vehicle's moving speed as input parameters for the computer (Fig 11).

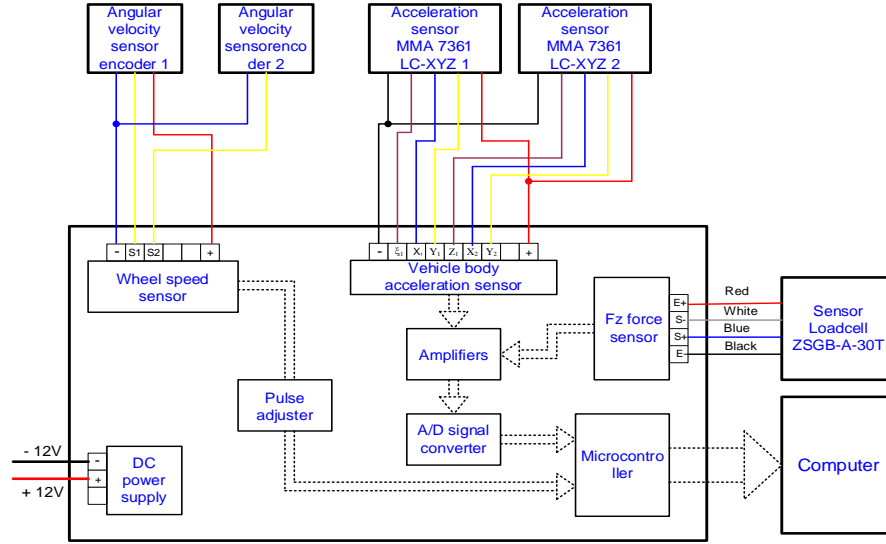

Figure 11. Arrangement diagram of sensors measuring signals and processing measurement results.

Install the sensors according to the experimental layout diagram. The experimental equipment is specialized dynamic testing equipment from Kistler (Fig 13), Germany

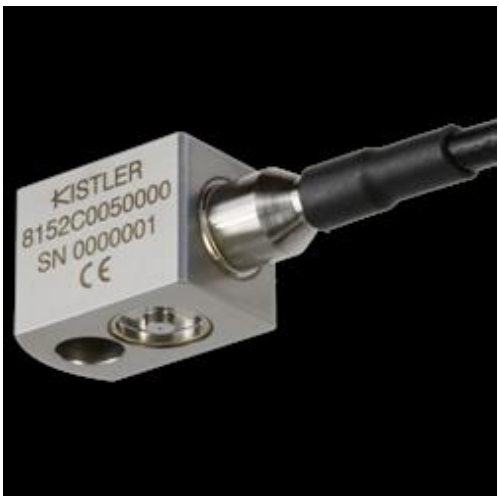 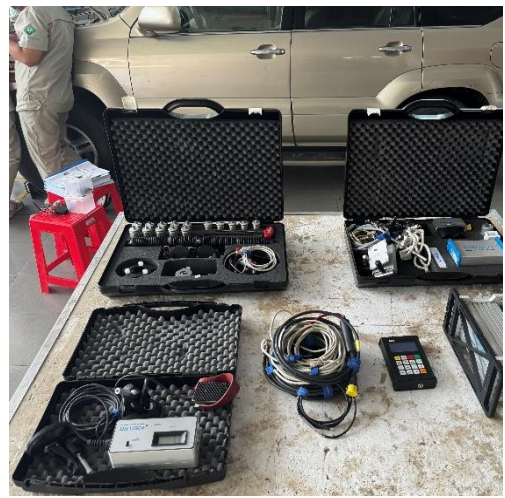

Figure 12. Acceleration sensor 8152C Kistler      **Figure 13.** Kisler dynamic experimental equipment

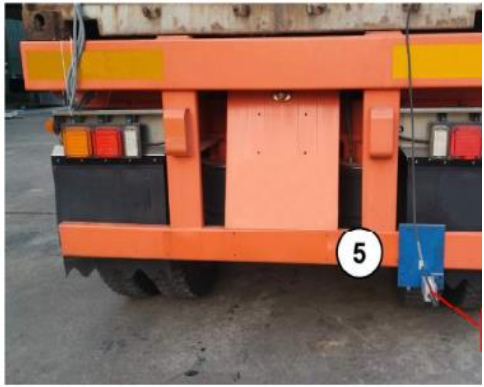
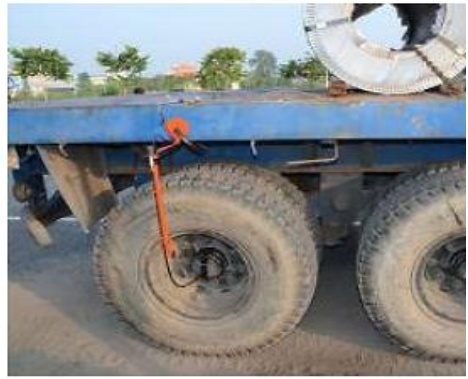

Figure 14. Install vehicle acceleration sensors and vehicle body displacement sensors on semi-trailers

Figure 15. Install the wheel speed sensor

Figure 14, and Figure 15 shows that the vehicle body acceleration sensor, wheel speed sensor, and load cell force sensor are installed on the vehicle at appropriate locations to collect data and send it to the computer.

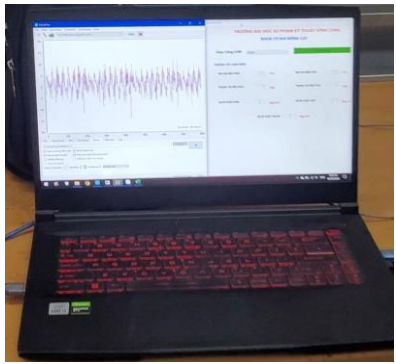
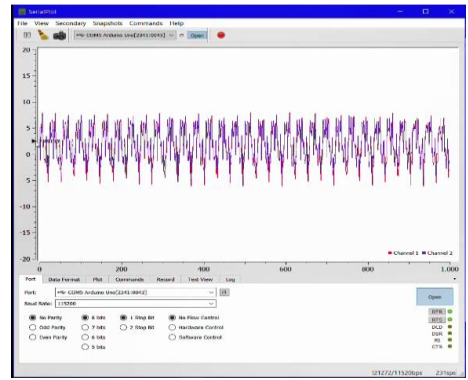

Figure 16. The computer receives and processes measured data from the sensors

Figure 17. Displays the values collected from the experiment

The computer receives and processes signals sent from sensors and calculates (Figure 16) and displays the vehicle body acceleration value and dynamic load Fz through the display screen (Figure 17).

To compare and verify theory with experiment, the experiment was performed with the same input parameters as the theoretical simulation calculation conditions with a travel speed of 50 km/h when the vehicle is full. Load, the driver holds the accelerator pedal to keep the car moving continuously on road profile C according to ISO 8608:2016 [2] standards.

Experimental results determining the vertical reaction force from the road surface to the wheels on the bridges, thereby determining the largest dynamic load coefficient, show that $k_{đmax}$ when the vehicle is running at a speed of 50km/h, the largest dynamic load coefficient is

k<sub>đmax</sub> = 1.38; The average maximum acceleration of the vehicle body $\ddot{Z}_{a\max}$ when experimentally obtained the value $\ddot{Z}_{a\max\ x}$= 2.52. The average experimental results of vehicle body acceleration when compared with simulation theory have a maximum deviation of 7.4%; The deviation of the maximum dynamic load coefficient k<sub>đmax</sub> is 8.1%.

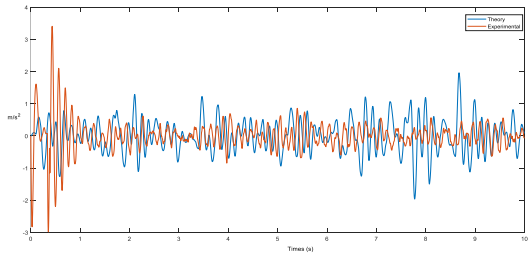

Figure 18. Compare the average acceleration of the vehicle body between theory and experiment

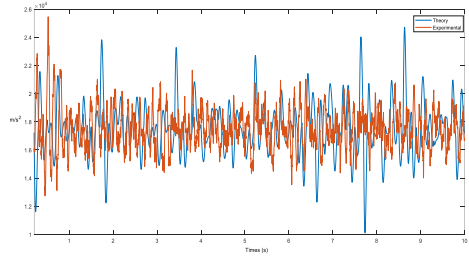

Figure 19. Compare the vertical reaction force acting from the road surface on the wheels at Bridge 1

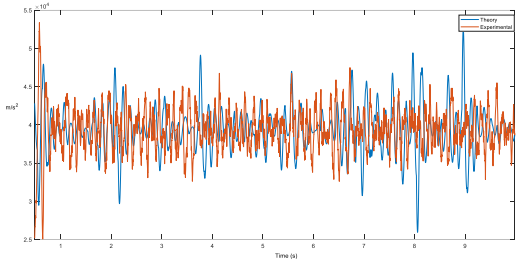

Figure 20. Compare the vertical reaction force acting from the road surface on the wheels at Bridge 2

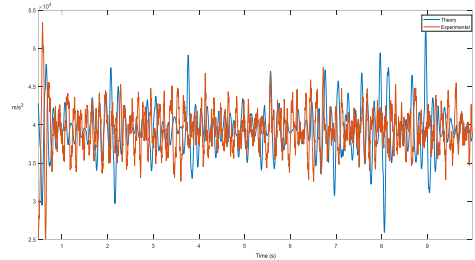

Figure 21. Compare the vertical reaction force acting from the road surface on the wheels at Bridge 3

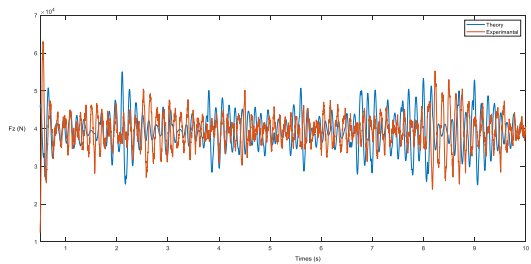

**Figure 22.** Compare the vertical reaction force acting from the road surface on the wheels at Bridge 4

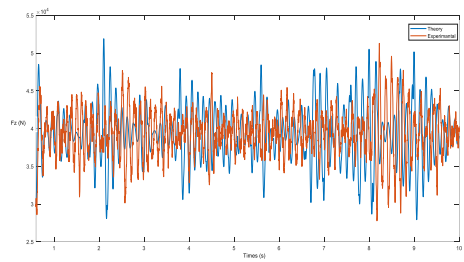

**Figure 23.** Compare the vertical reaction force acting from the road surface on the wheels at Bridge 5

Figures 18, 19, 20, 21, 22, and 23 show the experimental results of determining the average acceleration values of the vehicle body and the vertical reaction force of the road surface acting on the wheels of bridges 1, 2, 3. , 4, 5 and compare with the theoretical simulation value.

Table 1. Comparing values between theoretical simulation and experiment

|  | Average vehicle body acceleration $\ddot{Z}_{a\max}$ (m/s²) | Dynamic load coefficient $k_{đmax}$ |
|---|---|---|
| Theory | 2.36 | 1.25 |
| Experiment | 2.52 | 1.38 |
| Deviation | 7.4 % | 8.1 % |

When comparing simulation and experiment, the statistics on average vehicle body acceleration $\ddot{Z}_{a\max}$ and dynamic load coefficient $k_{đmax}$ in Table 1 indicate that the change values according to theoretical and experimental calculations are the same. By deducting the small value from the large value and dividing the result by the large value, the greatest difference between the theory and the experiment may be found.

The average vehicle body acceleration deviation is 7.4% and the dynamic load coefficient $k_{đmax}$ is 8.1% when comparing the results of the experiment and theory. This variation is acceptable given the assumptions used to develop the simulation model. The findings indicate that the vehicle dynamic model reliably and accurately captures the physical processes occurring in the system, with a discrepancy between the survey results using the theoretical model and the experimental results that are not too great.

4. Conclusion

To calculate the average acceleration of the vehicle's body and the dynamic load car, the research piece constructed a model, wrote a dynamic equation, and chose suitable plans, tools, subjects, and experimental techniques with Vietnam's current state of roads. Accurate and trustworthy test results are produced by measuring car body acceleration and dynamic stresses using German Kistler sensors.

Through theoretical research results, when evaluated by the vibration smoothness index, with a speed of 50 km/h, the vehicle moves most smoothly (average vertical acceleration does not increase suddenly) on road surfaces C. and D $\ddot{Z}_{a\max}$ < 3m/s² (still within the safety limit for goods) [6], [7]; At road surfaces E and F, the average acceleration of the vehicle body is $\ddot{Z}_{a\max}$ = [7.2; 12.4] m/s² > 3 m/s² [6], [7] (exceeds the warning limit, vehicle or road repair planning threshold), drivers need to choose a speed lower than 50 km/h when traveling to ensure Safety for goods, vehicles and roadbeds. When evaluated by dynamic safety criteria, the dynamic load coefficient $k_{đmax}$ when the vehicle moves on road sections C and D are within the limit to ensure the durability of vehicle details $k_{đmax}$ < 1.5 [9]. For road types E and F, the dynamic load value $k_{đmax}$ respectively $k_{đmax}$ exceeds the warning limit on the durability of car parts $k_{đmax}$ > 1.5 [9]. With a travel speed of 50 km/h, the two sides of roads C and D still ensure dynamic safety and vehicle longevity; The two routes E and F do not choose a speed of 50 km/h when traveling because it will exceed the kinematic safety warning limit and the durability of the car's parts.

Research on the law of value variation reveals comparable results when determining dynamic loads through theoretical and experimental methods. The vehicle's dynamic safety is reduced while driving on poorer road profiles because dynamic load values rise as a result of an increase in the dynamic load coefficient's maximum value. For this reason, when driving, it is important to select the proper speed for each type of road.

By utilizing the experimental arrangement diagram mentioned above, one may ascertain the mean vehicle body acceleration and dynamic load coefficient ($k_{đmax}$) and do a comparative analysis to validate the accuracy of the theoretical model. The difference between the outcomes of the theory and experiment, $\ddot{Z}a\max$ and $k_{đmax}$, is less than 10% [7.4%; 8.1%]. The aforementioned testing method's drawback is that it necessitates extensive planning and numerous experiments to produce the most accurate findings; yet, it is the experimental approach most suited to Vietnamese road conditions.

References

[1] Algirdas anulevicius, Kazimieras Giedra (2002) The evaluation of braking efficiency of tractor transport aggregate. Department of Transport and Power machinery, Lithuanian University of Agriculture

[2] ISO 8608:2016, Measurement and evaluation of mechanical vibration and shock as applied to machines, vehicles and structures.

[3] Nguyen Thanh Tung, (2017) "Research on braking effectiveness on roads with different adhesion coefficients of semi-trailers as a basis for proposing solutions to reduce traffic accidents", Doctoral thesis in engineering, Hanoi Polytechnic University.

[4] Mitschke, M., (1990), Dynamik der Kraftfahrzeuge, Band C: Fahrverhalten. Berlin: Springer.

[5] Mitschke, M., (1992), Dynamik der Kraftfahrzeuge, Band B: Fahrverhalten. Berlin: Springer.

[6] Mitschke, M., (1995), Dynamik der Kraftfahrzeuge. Springer Verlag, 3rd edition.

[7] Mitschke, M., (1986), Einfluss der Strasseenunebenheit auf Fahrzeugschwingung, IFF - Bericht, TU Braunschweig.

[8] Phan Tuan Kiet (2018), "Research on determining the vertical dynamic load of a vehicle fleet on the road surface", Doctoral thesis in engineering, Hanoi University of Science and Technology.

[9] Vo Van Huong (2014), "Automotive dynamics", Vietnam Education Publishing House, pages 74-78.

[10] Le Van Quynh, Zhang Jianrun, Lui Xiaobo, Wang Yuan (2011), "Nonlinear dynamic analysis of interaction between vehicle and road surface for 5-axle heavy truck", Journal of Southeast University (English Edition) Vol.27, No.4, pp.405-409, doi:10.3969/j.issn.1003-7985.2011.04.012.

[11] Vanliem Nguyen, Renqiang Jiao, Jianrun Zhang (2020) "Control Performance of Damping and Air Spring of Heavy Truck Air Suspension System with Optimal Fuzzy Control", SAE Int. J. Veh. Dyn., Stab., and NVH 4(2):179-194, 2020.

[12] TCVN 2737-1995, Loads and impacts. Institute of Construction Science - Ministry of Construction compiled, Department of Science and Technology - Ministry of Construction proposed, Ministry of Construction issued.

[13] Rohani RBMM, Abdullah ME (2018) Dynamic Load coefficient of Tyre forces from truck axles. Appl Mech Mater 405–408(2013):1900–1911.

[14] Wang J (2012) Nonlinear modeling and h-infinity model reference control of pneumatic suspension system. Ph.D thesis, Iowa State University, Canada.

[15] Moheyeldein MM, Abd-El-Tawwab AM, Abd El-gwwad KA, Salem MMM (2018) An analytical study of the performance indices of air spring suspensions over the passive suspension. Automotive and Tractors Eng. Depart., Faculty of Engineering, Minia University, El-Minia.

[16] Abid HJ, Chen J, Nassar AA (2015) Equivalent air spring suspension model for quarter-passive model of passenger vehicles. Int Sch Res Not. https://doi.org/10.1155/2015/974020.

[17] Tang G, Zhu H, Zhang Y, Sun Y (2015) Studies of air spring mathematical model and its performance in cab suspension system of commercial vehicle. SAE International by Columbia Univ.

[18] Hondo T, Tanaka T (2020) Investigation of relationship between initial setting of leveling valves and air spring pressure of a railway vehicle when assuming the centrifugal force action. In: IAVSD 2019: Advances in dynamics of vehicles on roads and tracks. Springer Nature Switzerland AG, pp 252–260 24. Razdan S.